\input{aipcheck}

\documentclass[
    ,final            
  ]
  {aipproc}

\layoutstyle{6x9}

\newcommand{\ima}{\hbox{Im}\,}
\newcommand{\rea}{\hbox{Re}\,}

\begin{document}

\title{Dispersive chiral approach to Meson-meson dynamics:
Spectroscopy results for light scalars and precision studies
\footnote{Invited Plenary talk at the Conference
``Quark Confinement and the Hadron SpectrumVII''. 2-7th Sept. 2006.
Ponta Delgada, Açores. Portugal. }}


\author{J. R. Pel\'aez}{
  address={Departamento de F\'{\i}sica Te\'orica II, 
Universidad Complutense, 28040 Madrid, SPAIN }
}

\begin{abstract}
Dispersive approaches provide model 
independent description of meson-meson scattering. 
We first review here 
the use of dispersion relations to obtain a model independent unitarization
of Chiral Perturbation Theory amplitudes, 
that establish the existence of light scalar mesons and
whose leading $1/N_c$ behavior
suggest they ahve a non $\bar qq$ dominant component.
We also review the forward dispersion relation 
checks on conflicting experimental data and the
resulting  very precise $\pi\pi$ scattering scattering amplitudes. 
\end{abstract}

\maketitle


Over the last years there has been a renewed interest in meson-meson scattering.
The reasons are that its low energy description 
is relevant for understanding the QCD vacuum,
and that the description below 1.2 GeV requires
the existence of some poles in the amplitudes, related to 
light scalar mesons. Such resonances are the subject of a strong debate,
first on their existence, that seems to be settling, but also on their nature.

Data on meson-meson scattering is obtained indirectly from other processes
involving nucleons or the decays of other heavy mesons. 
Thus, very often the existing data on  meson-meson amplitudes are extracted with 
strong model dependent assumptions (one-pion exchange, absorption models, 
rescattering, pole extrapolations to define the initial meson-meson state, etc...), 
and are therefore plagued with
systematic uncertainties much larger than statistical errors.
There are also $K\rightarrow\pi\pi l \nu$ experiments, known as $K_{l4}$ decays,
where the  pion phase shifts are extracted in a particular combination
of isospin 0 and 1, free of the previous
systematic uncertainties and yield very precise and reliable
results, but limited to invariant masses smaller than the kaon
mass. 

Concerning the theoretical description, we have two model independent
approaches. On the one hand, we have  Chiral Perturbation Theory (ChPT), which is 
the effective Lagrangian of QCD, written as an expansion in masses and derivatives
of pions, kaons and etas, which are the Goldstone bosons 
of the spontaneous chiral symmetry breaking of QCD. The only caveat to this 
systematic expansion is that it can only be applied at low energies.
On the other hand, it is possible to use the usual S-matrix 
constraints of causality, analyticity,  unitarity, crossing, etc...
 to write dispersion relations for the different meson-meson channels.
We review here how we have recently applied this approach to check 
the consistency of different data sets and to obtain a precise 
pion-pion scattering amplitudes in the whole energy range.
Of course, both approaches can also be combined to obtain a 
model independent description of data that also incorporates the low energy
chiral symmetry constraints (see. H. Leutwyler's talk on Roy equations 
in this conference). 
We review here how this has been done by means of the Inverse Amplitude Method (IAM), 
which has the advantage that only uses
ChPT input in the dispersive integrals and therefore
allows for a relation between the resulting fits and QCD. Since the IAM generates
the light resonances that appear in scattering, it is then possible to
study their nature in terms of QCD parameters, like the number of colors.

\vspace*{-.3cm}
\section{The Inverse Amplitude Method from Dispersion Theory}

We will review here how the {\it one-channel} Inverse Amplitude Method (IAM) 
\cite{Truong:1988zp,Dobado:1992ha,Dobado:1996ps}
for pion-pion scattering is obtained just by using ChPT up to a given order inside 
a dispersion relation. There are no further assumptions and 
therefore the approach is model independent and provides an elastic
amplitude that satisfies unitarity and has the correct ChPT 
expansion up to that given order. 

To fix ideas, let us consider the ChPT series for a pion-pion scattering 
partial wave amplitude of definite isospin $I$ and angular momentum $J$, namely,
$t_{IJ}=t_{IJ}^{(2)}+t_{IJ}^{(4)}+...$ 
where $t_2=O(p^2)$, $t_4=O(p^4)$ and $p$ stands for the 
pion mass or momentum. For the complete partial wave $t_{IJ}(s)$, it is possible to write
a dispersion relation 
\begin{equation}
t_{IJ}(s)=C_0+C_1s+C_2s^2+
\frac{s^3}\pi\int_{s_{th}}^{\infty}\frac{\ima
t_{IJ}(s')ds'}{s'^3(s'-s-i\epsilon)} + LC(t_{IJ}),
\label{disp}
\end{equation}
that, for convenience, we have subtracted three times. Note we have explicitly
written the integral over the right hand cut (or physical cut, 
extending from threshold, $s_{th}$ to infinity) 
but we have abbreviated by $LC$ the equivalent
expression for the left cut (from 0 to $-\infty$). We could do similarly with 
other cuts, if present, as in the $\pi K$ case.

We can also write dispersion relations for $t^{(2)}$ and $t^{(4)}$, 
but remembering that $t^{(2)}$ is a pure tree level amplitude and 
it does not have imaginary part nor cuts:
\begin{eqnarray}
t_{IJ}^{(2)} = a_0+a_1s, \qquad  
t_{IJ}^{(4)} = b_0+b_1s+b_2s^2+     
\frac{s^3}\pi\int_{s_{th}}^{\infty}\frac{\ima
t_{IJ}^{(4)}(s')ds'}{s'^3(s'-s-i\epsilon)}+LC(t^{(4)}_{IJ}).
\label{disp1}
\end{eqnarray}
We now recall that unitarity, for physical values of $s$ in the elastic region implies:
\begin{equation}
  \ima t_{IJ} =\sigma \vert t_{IJ}\vert^2 \quad\Rightarrow \quad\ima \frac{1}{t_{IJ}}=-\sigma \quad\Rightarrow\quad
t_{IJ}=\frac{1}{\rea t_{IJ}^{-1} - i \sigma},
\label{unit}
\end{equation}
where $\sigma=2p/\sqrt{s}$.  Therefore,
the imaginary part of the {\it inverse amplitude} is {\it exactly} known in the elastic regime.
We can then write a dispersion relation like that in \eqref{disp} 
but now for the auxiliary function $G=(t_{IJ}^{(2)})^2/t_{IJ}$, i.e.,
\begin{equation}
G(s)=G_0+G_1s+G_2s^2+     \\   \nonumber
\frac{s^3}\pi\int_{s_{th}}^{\infty}
\frac{\ima G(s')ds'}{s'^3(s'-s-i\epsilon)}+LC(G)+PC,
\label{Gdisp}
\end{equation}
where now $PC$ stands for possible pole contributions in $G$ coming from 
zeros in $t_{IJ}$. It is now straightforward to expand the subtraction constants
and use that
 $\ima t_{IJ}^{(2)}=0$ and  $\ima t_{IJ}^{(4)}=\sigma\vert t_{IJ}^{(2)}\vert^2$,
so that $\ima G= -\ima t_{IJ}^{(4)}$. In addition,
 up to the given order, $LC(G)\simeq -LC(t_{IJ}^{(4)})$,
whereas $PC$ is of higher order and can be neglected. Thus
\begin{eqnarray}
\frac{t_{IJ}^{(2)2}}{t_{IJ}}\simeq a_0+a_1s-b_0-b_1s-b_2s^2 
-\frac{s^3}\pi\int_{s_{th}}^{\infty}\frac{\ima
t_{IJ}^{(4)}(s')ds'}{s'^3(s'-s-i\epsilon)}-LC(t_{IJ}^{(4)})
\simeq t_{IJ}^{(2)}-t_{IJ}^{(4)}.
\label{preIAM}
\end{eqnarray}
We have thus arrived to the so-called Inverse Amplitude Method (IAM):
\begin{equation}
t_{IJ}\simeq
t_{IJ}^{(2)2}/(
t_{IJ}^{(2)}-t_{IJ}^{(4)} ),
\label{IAM}
\end{equation}
that provides an elastic amplitude that satisfies unitarity and has the correct
low energy expansion of ChPT up to the order we have used. It is straightforward
to extend it to other elastic channels or to higher orders \cite{Dobado:1996ps}.
Note also that, by looking at \eqref{unit}, it seems that it can also
be derived by replacing $\rea t_{IJ}^{(-1)}$ by its $O(p^4)$ ChPT expansion.
But, strictly speaking,
 \eqref{unit} is only valid in the real axis, whereas our derivation allows
us to consider the amplitude in the complex plane, and, 
in particular, look for poles of the associated resonances. Actually,
already ten years ago \cite{Dobado:1996ps}, 
with the single  channel IAM we were able to 
generate poles for the $\rho(770)$, $K^*(892)$ and most interestingly, 
the controversial $\sigma$ (also called $f_0(600)$), 
{\it without any model dependent assumptions}.

Note that {\bf the above one-channel IAM derivation is 
 model independent},
and that contrary to a wide belief in the community {\bf contains a left cut}
and {\it respects crossing symmetry up to}, of course, {the order in the ChPT
expansion that has been used}.

The confusion may come from the fact that the IAM
has also been applied in a coupled channel formalism,
for which {\it there is still no dispersive derivation},
and sometimes with further approximations.
Indeed one can arrive to \eqref{IAM} in a matrix form, ensuring
coupled channel unitarity, just by expanding
the real part of the inverse T matrix. 
{\it For the coupled channel case} different approximations to $Re T^ {-1}$
have been used:

$\bullet$ The fully renormalized one-loop ChPT calculation of $Re T^ {-1}$ 
provides the correct ChPT expansion in all channels,
also with left cuts approximated to $O(p^4)$ \cite{Guerrero:1998ei,GomezNicola:2001as}. 
Indeed, using
ChPT parameters consistent with previous determinations within standard ChPT,
it was possible \cite{GomezNicola:2001as,Pelaez:2004xp}
to describe below 1.2 GeV all the scattering
channels of two body states made of pions, kaons or etas. Simultaneously,
this approach \cite{Pelaez:2004xp} generates poles associated to the $\rho(770)$, $K^*(892)$ vector mesons,
and the  $f_0(980)$, $a_0(980)$, $\sigma$ 
and $\kappa$ (also called $K_0(800)$) scalar resonances.

$\bullet$ Originally \cite{Oller:1997ng}, the coupled channel
IAM  was used neglecting the crossed loops and tadpoles.
This approach is considerably 
simpler, and although
 it is true that the left cut is absent, its numerical influence
was shown to be rather small, since the meson-meson data
are nicely described with very reasonable
chiral parameters and generates all the poles enumerated above. 
Let us remark that this approximation keeps the s-channel loops
but also the tree level up to $O(p^4)$, and that this tree level
encodes the effect of  heavier resonances, like the rho. Thus,
contrary to some common belief, this approach still incorporates,
for instance,
the low energy effects of t-channel rho exchange.

$\bullet$ Finally, if one is interested in describing 
just the scalar meson-meson channels, it is possible to 
use just one cutoff (or even a dimensional regularization scale)
that numerically mimics 
the combination of chiral parameters that appear in those 
scalar channels. This has become very popular, 
even beyond the meson-meson interaction realm,
due to its great simplicity but remarkable success
\cite{Oller:1997ti}.

\vspace*{-.3cm}
\section{Nature of light scalars from Unitarized ChPT}

One of the big advantages of the unitarization approaches described 
in the previous sections is that when
they use the fully renormalized ChPT amplitudes,
they therefore have the correct chiral and flavor symmetry structure,
including both the spontaneous and explicit symmetry breaking. 
Furthermore
we also have the correct dependence 
on QCD parameters like the number of colors, which 
is of particular interest, since 
there are sharp  predictions on how the mass and width 
of $\bar qq$ resonances should behave in a large $N_c$ QCD expansion.
In particular, $M\simeq O(1)$ whereas $\Gamma\simeq O(1/N_c)$.

The $1/N_c$ leading behavior of all 
ChPT parameters is known and model independent, so that they can be
varied accordingly to study \cite{Pelaez:2003dy} the $N_c$ dependence 
of the amplitudes.
In particular, we can study the $N_c$ behavior of all the poles
generated in the IAM. It has been shown that both the mass and width
of the vector mesons generated
with the IAM follow remarkably well the expected $\bar qq$ behavior.
However, {\it light scalars do not follow a dominant $\bar qq$ $N_c$ behavior}
\cite{Pelaez:2003dy},
at least for $N_c$ not too far from real life, $N_c=3$.
In figure 1 we illustrate these two different behaviors calculated at
with the $O(p^4)$ IAM, for the $\rho$  $K^*(892)$, and for the $\sigma$,
and $\kappa$, whose poles can be obtained with the one-channel
IAM, and therefore in a model independent way.
Similar plots can be found for the other scalars in \cite{Pelaez:2003dy}.

At this point it is worth noting that two-meson loop diagrams are subdominant
at large $N_c$. Indeed, the above results imply that for vectors
the meson loops play a small role in the cancellations that lead to poles
in the amplitude, whereas for scalars such loop diagrams are very important
at $N_c=3$. Since these diagrams become smaller and smaller one could wonder about
the influence of higher order effects in scalars. Thus recently, we have performed
\cite{Pelaez:2006nj}
the full two-loop IAM analysis of the $\sigma$ channel confirming that,
as it happened for the $O(p^4)$ amplitude, 
close to $N_c=3$ the sigma behaves rather differently than expected for a $\bar qq$
state, but that as the loop diagrams are suppressed, a subdominant $\bar qq$
behavior is recovered at larger $N_c$. 
Remarkably, this $\bar qq$ behavior
arises slightly above 1 GeV, as it can be seen in Figure 1 (right plot). 
This seems to support the suggestion \cite{VanBeveren:1986ea} that
there is a non-$\bar qq$ scalar nonet below roughly 1 GeV, and another $\bar qq$
nonet above, but using a mode independent framework
based on dispersive integrals, ChPT and the large $N_c$ QCD behavior.

\begin{figure}
\includegraphics[height=.45\textwidth]{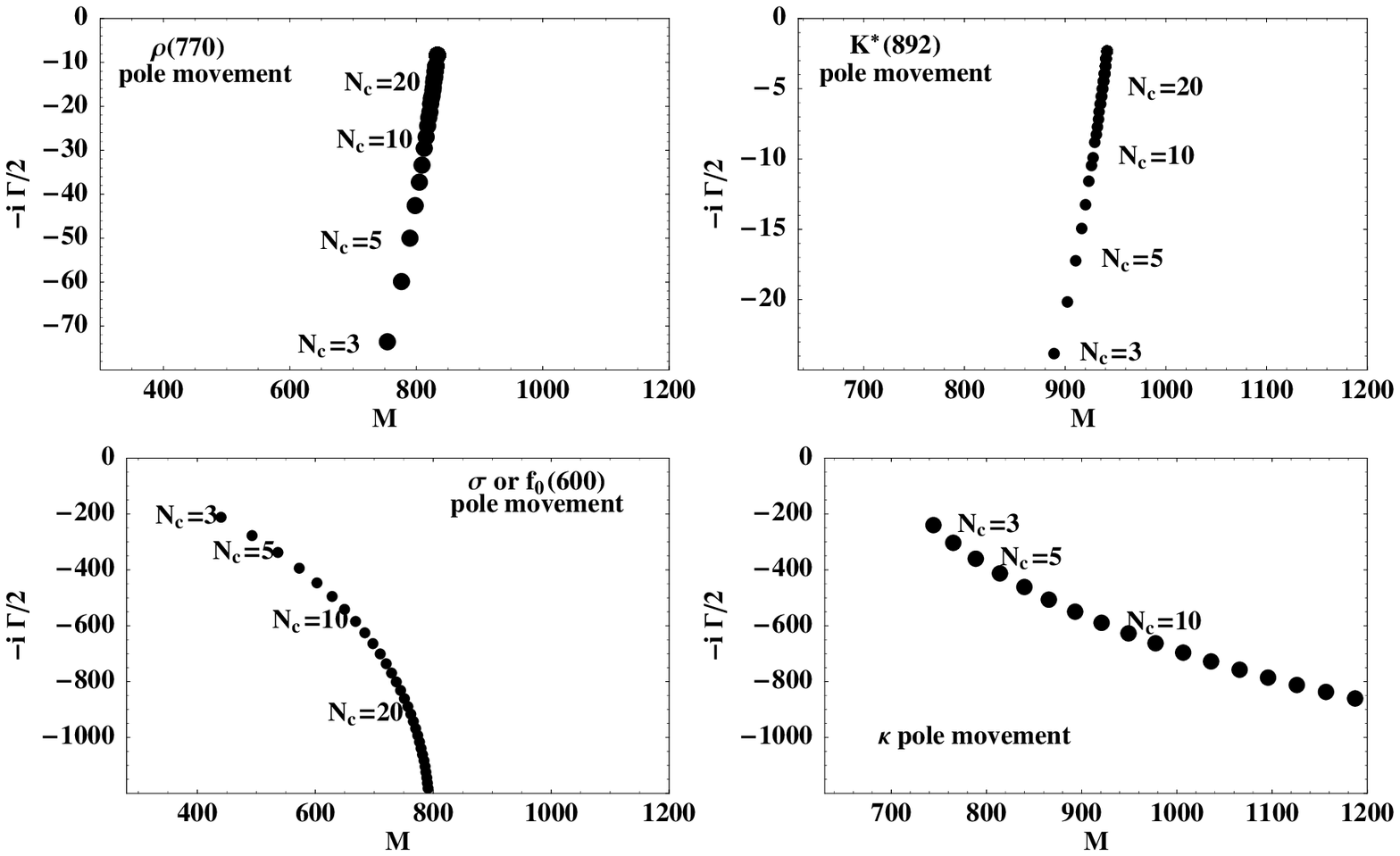}\hspace{-.2cm}
\includegraphics[height=.3\textwidth,angle=270,origin=cr]{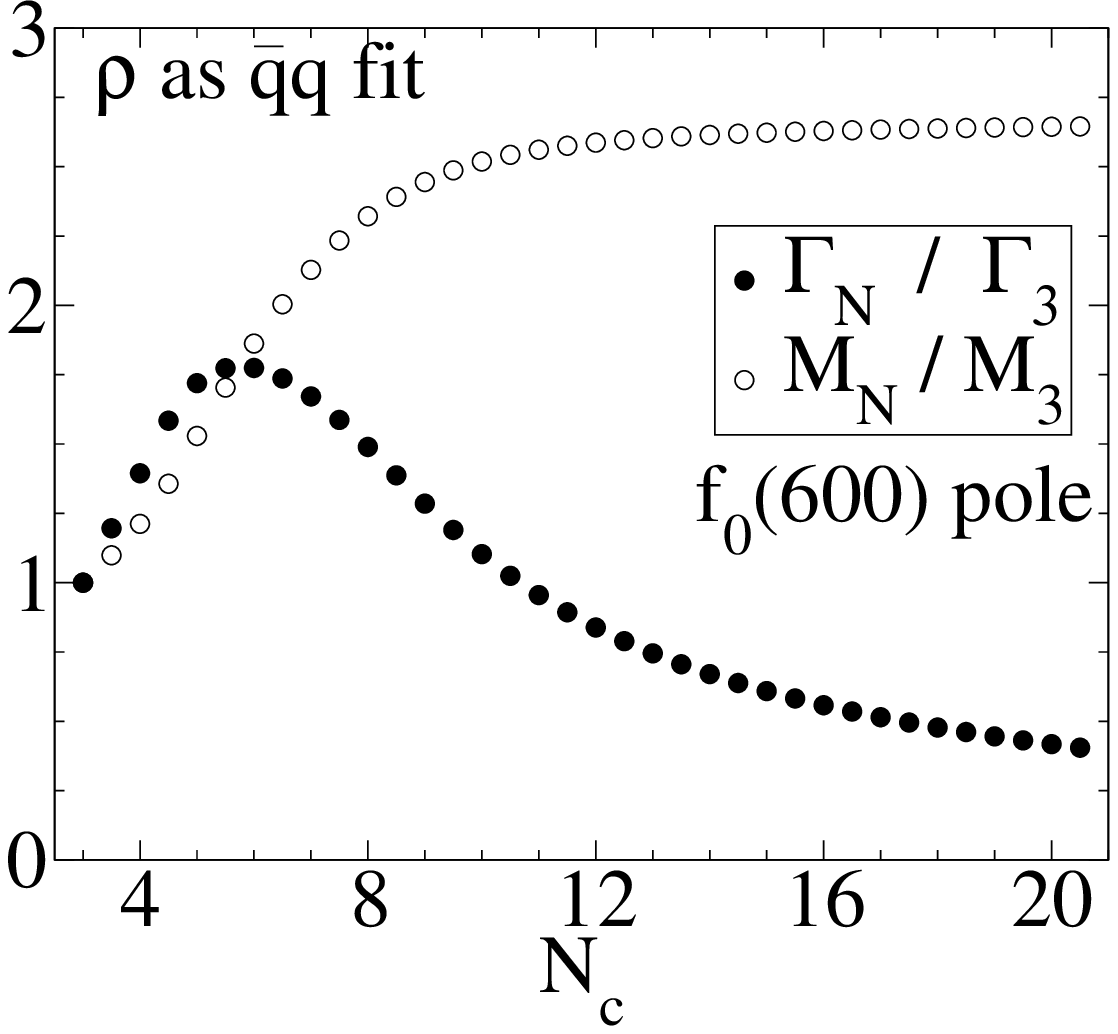}
  \caption{Leading $N_c$ behavior of resonance poles generated 
IAM. Left and center columns: with one-loop $O(p^4)$ ChPT
Both the $\rho$ and $K^*$ poles in the imaginary plane 
behave as expected for $\bar qq$ states, 
namely $M\simeq O(1)$ and $\Gamma\simeq O(1/N_c)$, respectively. In contrast, the
light scalars $\sigma$ and $\kappa$ do not behave predominantly as $\bar qq$.
Right: with two-loop $O(p^6)$ ChPT. We plot the mass $M$ and width $\Gamma$ 
evolution with $N_c$ calculated from the pole position, normalized to the $N_c=3$
case. Once more the dominant behavior is not that of a $\bar qq$ but a subdominant
$\bar qq$ behavior emerges at larger $N_c$ around 1 GeV in mass.}
\end{figure}

\vspace*{-.3cm}
\section{Precise amplitudes from Dispersion Relations and Roy equations}

In the previous approach we used ChPT inside the 
integrals of partial wave dispersion relations.
This has the advantage that we can control  all the parameters of our input,
and allow for their variation in order to understand the nature of
poles or the dependence on certain QCD parameters. However
ChPT does not necessarily give a good description of the integrands
at high energies. Thus, even though
we performed several subtractions to suppress the high energy regime
we cannot claim to have very precise amplitudes, but a
qualitative and most likely a semi-quantitative description.

There are however other approaches where one can use directly the data 
inside the integrals. Of course, now  it is much harder, if possible at all,
to interpret changes of parameters in terms of QCD,
but we can get extremely precise results for the amplitude and other observables like
scattering lengths, poles, etc... Recently,
ChPT constraints and data have been included in single channel dispersion 
relations for $\pi\pi$ and $\pi K$ 
{\it partial waves} \cite{Zhou:2006wm}, confirming the existence of poles for $\sigma$
and $\kappa$ resonances, introducing some cutoffs
on the dispersive integrals. 
For partial waves, the left cut 
is always a very delicate issue, because the large $t$ behavior is not well known. 
Two possible model independent 
approaches to this problem have been given in the literature: One is to rewrite 
the left cut integrals
in terms of a coupled set of integral equations relating different partial waves,
known as Roy equations for $\pi\pi$ scattering and Roy-Steininger equations for $\pi K$
scattering, which have been recently used to obtain precise determinations
of the $\sigma$ \cite{Caprini:2005zr} and $\kappa$ \cite{Descotes-Genon:2006uk} 
poles, respectively. 

Here I will comment on the other
approach, namely, to use  {\it Forward} Dispersion Relations (FDR),
that is, to avoid using partial waves and use full amplitudes setting $t=0$.
In a recent analysis we have used the following
set of dispersion relations that form a complete isospin set:
by choosing either $F=F_{00}$ or $F=F_{0+}$  in 
\begin{eqnarray}
\real F(s)-F(4M_{\pi}^2)=
\frac{s(s-4M_{\pi}^2)}{\pi}{\rm P.P.}\int_{4M_{\pi}^2}^\infty{\rm d} s'\,
\frac{(2s'-4M^2_\pi)\ima F(s')}{s'(s'-s)(s'-4M_{\pi}^2)(s'+s-4M_{\pi}^2)}.
\label{(4.1a)}
\end{eqnarray}
we have two dispersion relations which are {\it twice subtracted} .
Thus, the weight of the high energy part is quite suppressed, indeed as much
as in Roy equations, but with the advantage of having always {\it positive}
contributions to the integrand, a fact that makes them much more precise.
In addition, by setting $s=2M^2_\pi$, and  $F=F_{00}$,  we find
two sum rules important to fix the Adler zeros.
Finally, for the t-channel exchange of isospin 1, 
which does not require subtractions, we use, 
\begin{equation}
\real F^{(I_t=1)}(s,0)=\frac{2s-4M^2_\pi}{\pi}\,{\rm P.P.}\int_{4M^2_\pi}^\infty{\rm d} s'\,
\frac{\ima F^{(I_t=1)}(s',0)}{(s'-s)(s'+s-4M^2_\pi)}.
\label{(4.3)}
\end{equation}
At threshold this is known as the Olsson sum rule.

\addtolength{\floatsep}{-.8cm}
\begin{figure}
\includegraphics[width=.3\textwidth,angle=270]{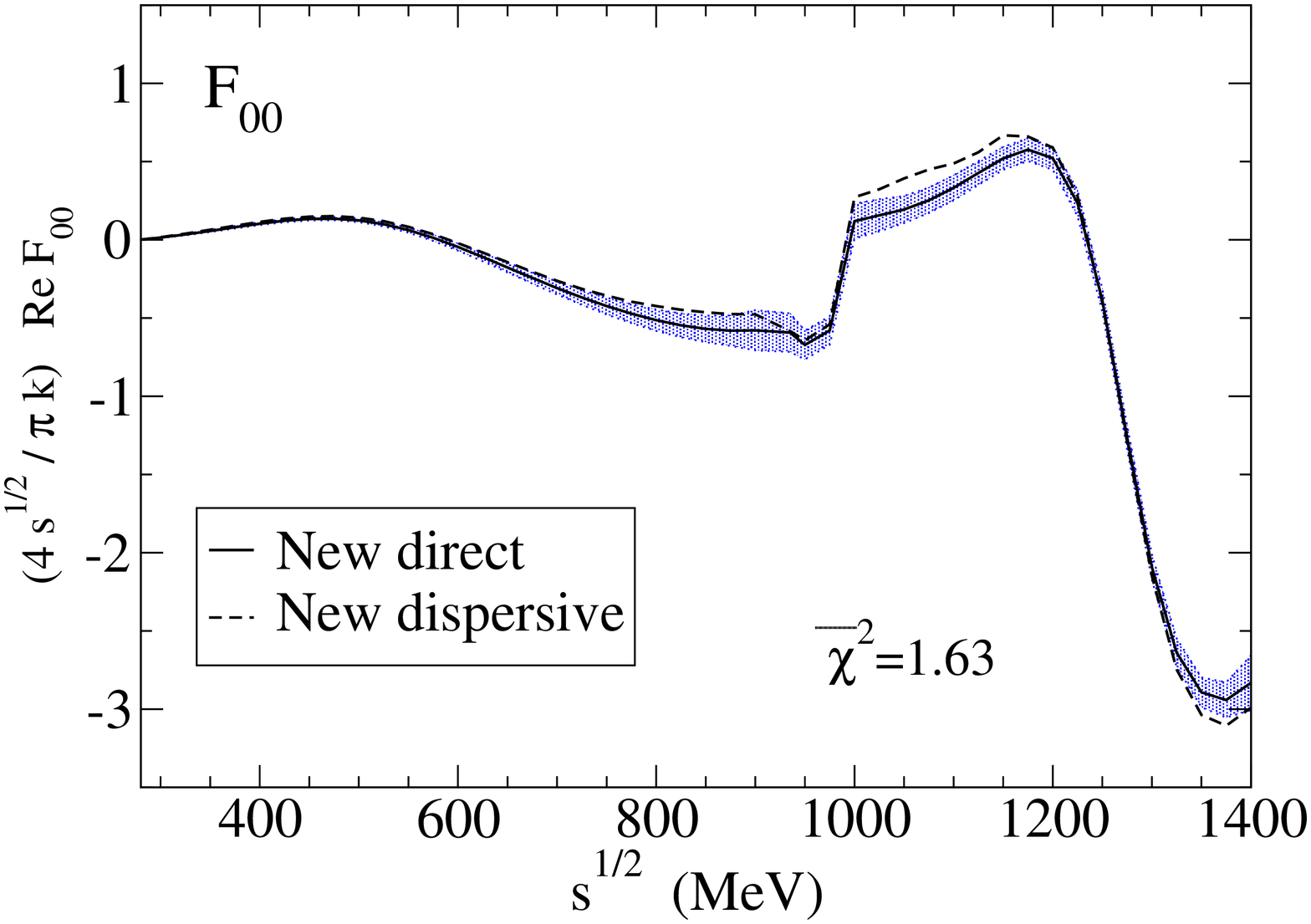}
\hspace{-.8cm}
\includegraphics[width=.3\textwidth,angle=270]{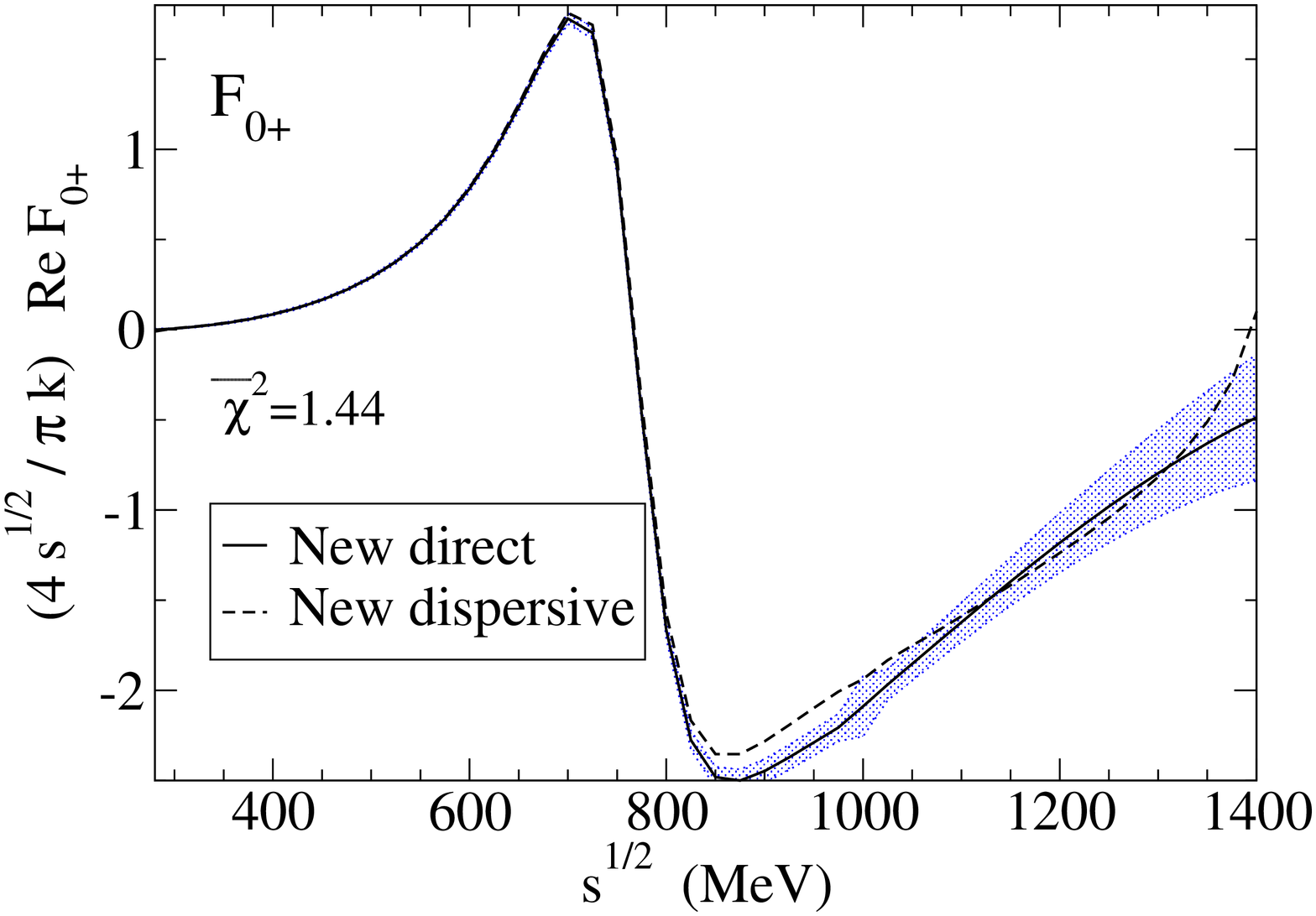}
\hspace{-.8cm}
\includegraphics[width=.3\textwidth,angle=270]{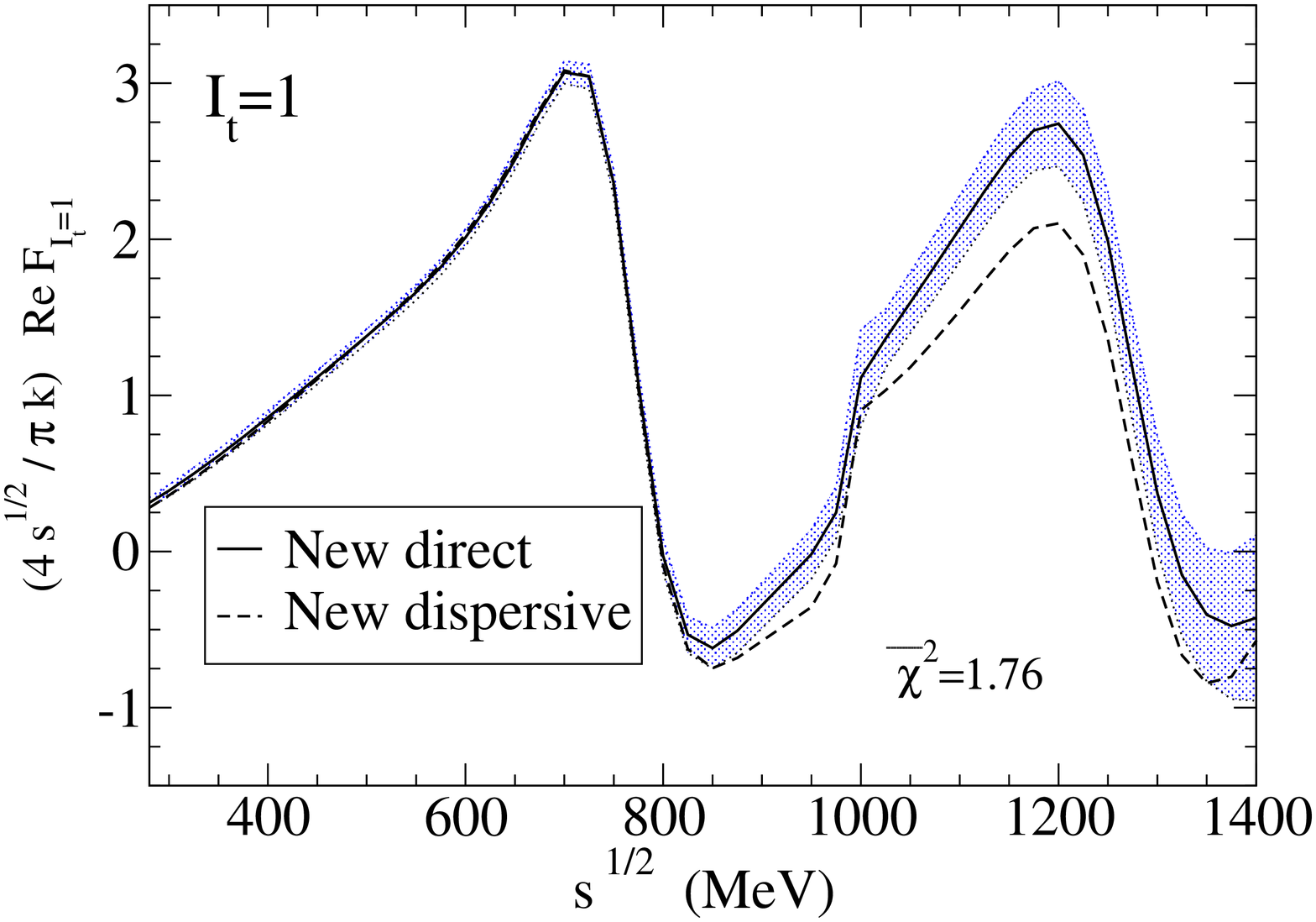}
\end{figure}
\begin{figure}
\includegraphics[width=.28\textwidth,angle=270]{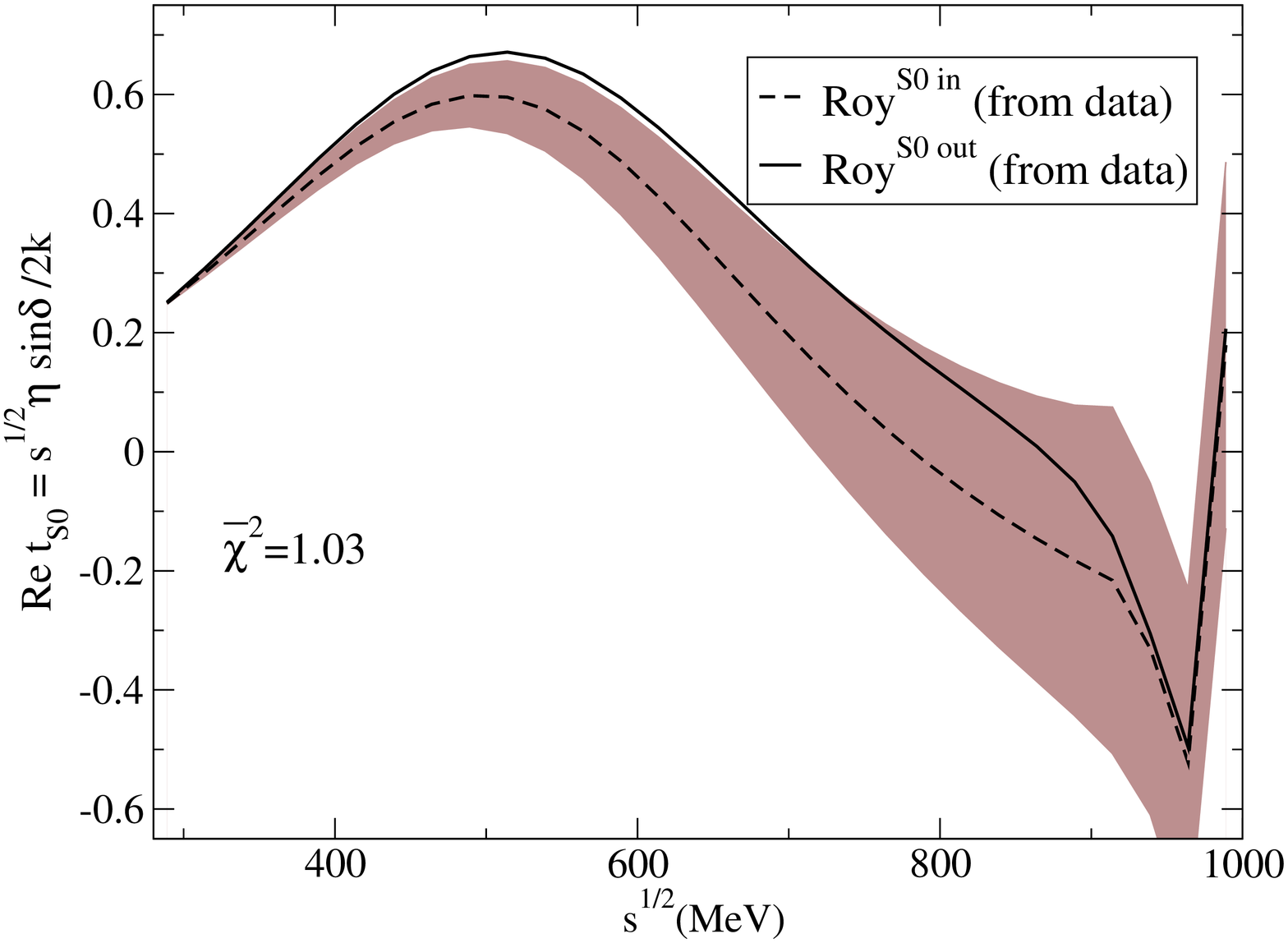}
\hspace{-.5cm}
\includegraphics[width=.28\textwidth,angle=270]{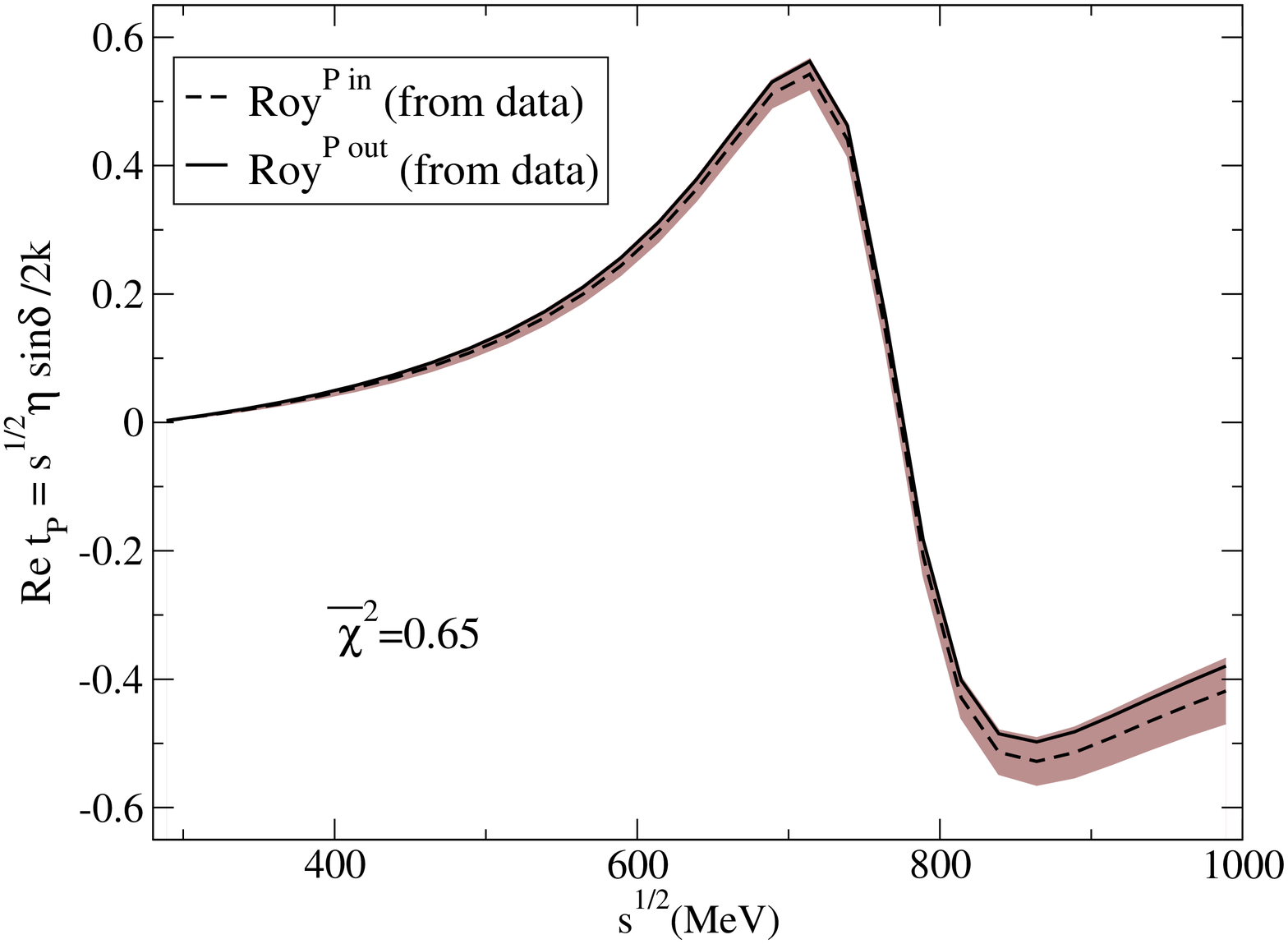}
\hspace{-.5cm}
\includegraphics[width=.28\textwidth,angle=270]{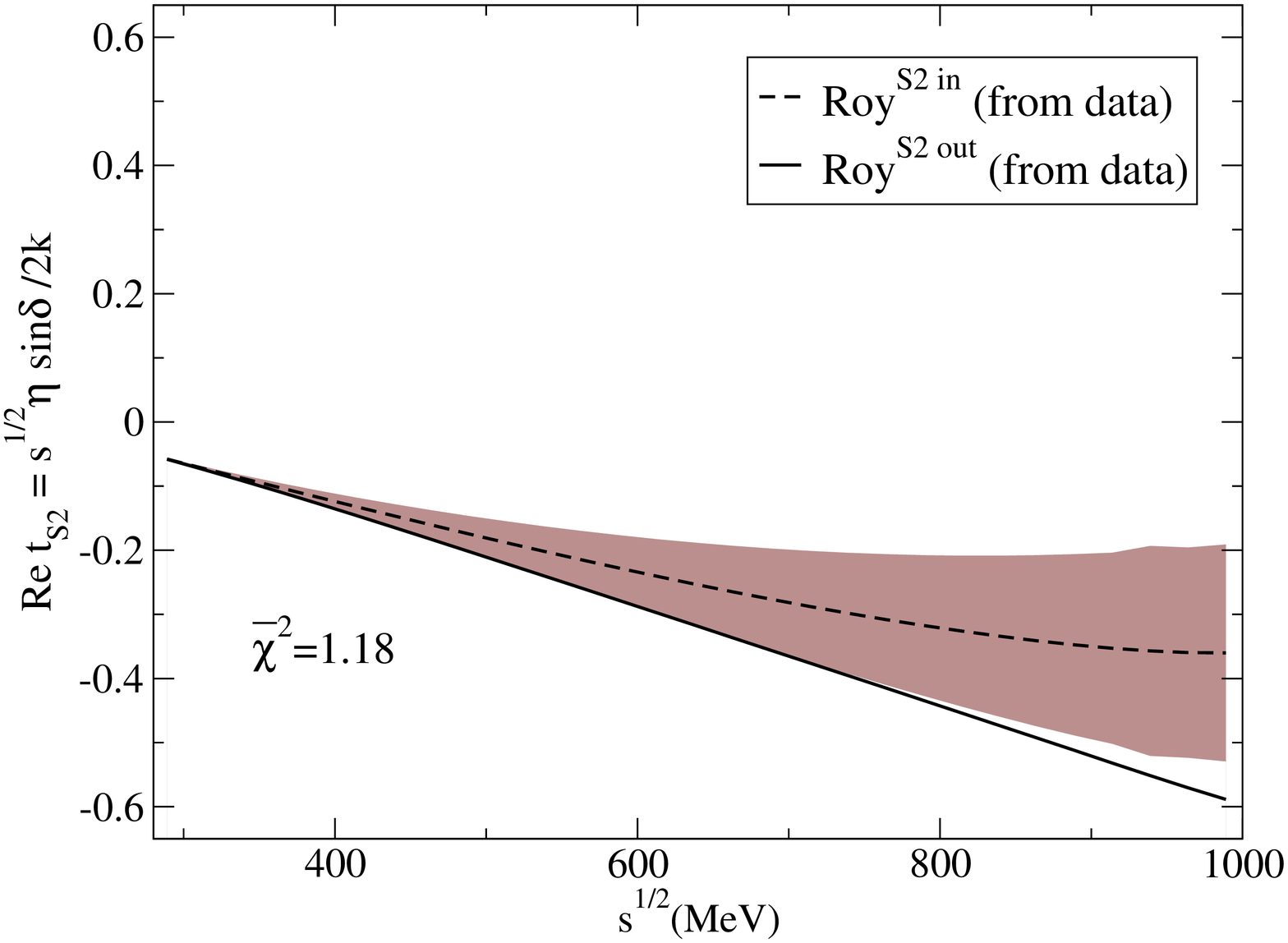}
  \caption{The data fits obtained in \cite{Kaminski:2006yv,Pelaez:2004vs}
satisfy remarkably well the complete set of FDR and Roy Equations below $KK$ threshold
and fairly well above. In this plots we compare the result of the direct parametrization
(direct or ``in'' curves) versus the integral representation (dispersive or ``out'').
We emphasize these curves come from fits to data, without FDR or Roy equations as constraints, so that the agreement within uncertainties 
between continuous and dashed lines is even more remarkable.
}
\end{figure}

First of all, we have found \cite{Pelaez:2004vs} that only a few 
sets of data available in the literature, satisfy reasonably
well the above forward dispersion relations within errors. Some of the most widely 
used experimental phase shift determinations 
indeed fail, on average for  each FDR, by more than 
2 standard standard deviations. The sets that satisfy FDR better
are those closer to the fit using just the low energy
data on  $K_{l4}$ decays commented in the introduction.
All these sets have a marked ``hunchback'' in the 400 to 900 MeV region
that undoubtedly is caused by the $\sigma$ pole. We show in Figure 2 (upper row)
the comparison between the FDR left hand sides (``direct'' calculation) and
FDR right hand sides (``dispersive'' calculation). Note that the agreement
below $KK$ threshold is astonishing given the tiny uncertainties and that
only data has been fitted, not the FDR themselves.
We have recently checked \cite{inprep} that our direct fits to data also satisfy
remarkably well Roy equations, as shown in Figure 2 (lower row).

In \cite{Pelaez:2004vs} we went further than this and improved the 
$\pi\pi$  amplitudes by constraining the fits to different partial waves to satisfy
the FDR below 950 MeV. In this way we obtained an even more precise representation 
that lies not too far from the best fits to data only, precisely 
those showing the $\sigma$ ``hunchback'' in the S0 wave.

Compared with the solution from Roy equations
in \cite{Colangelo:2001df}, the phase shift
solutions that satisfy better the FDR representation in \cite{Pelaez:2004vs}
agrees within errors in the low energy region for the S and P waves,
but the S0 phase shifts above 400 MeV lie about 2 standard deviations higher
than in \cite{Colangelo:2001df}. We also have a discrepancy of
about two standard deviations in the D wave scattering lengths and the
Regge residue of the $\rho$.

At present, and using our very recent improved fits 
above $KK$ threshold,
 we are working on an even more precise $\pi\pi$ amplitude
by constraining our fits to satisfy simultaneously FDR and Roy equations,
together with Froissart-Gribov and other sum rules. This
parametrization will be used to obtain a precise determination of
the $\sigma$ pole as well as of threshold parameters to
determine the low energy ChPT constants.

In summary, I believe that, dispersive approaches combined with ChPT
are the most powerful technique at our disposal to describe precisely
and understand in terms of QCD the meson-meson interactions and light scalar
spectroscopy.

\vspace*{-.3cm}
\begin{theacknowledgments}
\vspace*{-.3cm}
I thank the organization staff, and particularly E. Ribeiro,
for creating such a nice working environment
and offering me the chance to participate in this wonderful Conference.
\end{theacknowledgments}


\vspace*{-.3cm}
\begin{thebibliography}{9}
\vspace*{-.3cm}
\small

\bibitem{Truong:1988zp}
T.~N.~Truong,
Phys.\ Rev.\ Lett.\  {\bf 61} (1988) 2526.
  A.~Dobado, M.~J.~Herrero and T.~N.~Truong,
  Phys.\ Lett.\ B {\bf 235}, 134 (1990).

\bibitem{Dobado:1992ha}
  A.~Dobado and J.~R.~Pelaez,
  Phys.\ Rev.\ D {\bf 47}, 4883 (1993)


\bibitem{Dobado:1996ps}
  A.~Dobado and J.~R.~Pelaez,
  Phys.\ Rev.\ D {\bf 56}, 3057 (1997)

\bibitem{Guerrero:1998ei}
  F.~Guerrero and J.~A.~Oller,
  Nucl.\ Phys.\ B {\bf 537}, 459 (1999)
  [Erratum-ibid.\ B {\bf 602}, 641 (2001)]

\bibitem{GomezNicola:2001as}
  A.~Gomez Nicola and J.~R.~Pelaez,
  Phys.\ Rev.\ D {\bf 65}, 054009 (2002)


\bibitem{Pelaez:2004xp}
  J.~R.~Pelaez,
  Mod.\ Phys.\ Lett.\ A {\bf 19}, 2879 (2004)


\bibitem{Oller:1997ng}
  J.~A.~Oller, E.~Oset and J.~R.~Pelaez,
  Phys.\ Rev.\ Lett.\  {\bf 80}, 3452 (1998).
  Phys.\ Rev.\ D {\bf 59}, 074001 (1999)
  [Erratum-ibid.\ D {\bf 60}, 099906 (1999)]

\bibitem{Oller:1997ti}
  J.~A.~Oller and E.~Oset,
  Nucl.\ Phys.\ A {\bf 620}, 438 (1997)
  [Erratum-ibid.\ A {\bf 652}, 407 (1999)]

\bibitem{Pelaez:2003dy}
  J.~R.~Pelaez,
  Phys.\ Rev.\ Lett.\  {\bf 92}, 102001 (2004)

\bibitem{Pelaez:2006nj}
  J.~R.~Pelaez and G.~Rios,
  arXiv:hep-ph/0610397. To appear in Phys.\ Rev.\ Lett.

\bibitem{VanBeveren:1986ea}
  E.~Van Beveren, {\it et al.} 
  Z.\ Phys.\ C {\bf 30}, 615 (1986)
and 
hep-ph/0606022.
E.~van Beveren and G.~Rupp,
Eur.\ Phys.\ J.\ C {\bf 22} (2001) 493,
J.~A.~Oller and E.~Oset,
Phys.\ Rev.\ D {\bf 60} (1999) 074023.
  F.~E.~Close and N.~A.~Tornqvist,
  J.\ Phys.\ G {\bf 28}, R249 (2002)
(See. N.A. Tornqvist talk in this workshop)

\bibitem{Zhou:2006wm}
  Z.~Y.~Zhou and H.~Q.~Zheng,
  Nucl.\ Phys.\ A {\bf 775}, 212 (2006). 
  Z.~Y.~Zhou, {\it et al.},
  JHEP {\bf 0502}, 043 (2005)

\bibitem{Caprini:2005zr}
  I.~Caprini, G.~Colangelo and H.~Leutwyler,
  Phys.\ Rev.\ Lett.\  {\bf 96}, 132001 (2006)
(See. H. Leutwyler's talk in this conference)

\bibitem{Descotes-Genon:2006uk}
  S.~Descotes-Genon and B.~Moussallam,
  arXiv:hep-ph/0607133.


\bibitem{Pelaez:2004vs}
  J.~R.~Pelaez and F.~J.~Yndurain,
  Phys.\ Rev.\ D {\bf 71}, 074016 (2005)

\bibitem{Kaminski:2006yv}
  R.~Kaminski, J.~R.~Pelaez and F.~J.~Yndurain,
  Phys.\ Rev.\ D {\bf 74}, 014001 (2006)
  [Erratum-ibid.\ D {\bf 74}, 079903 (2006)]

\bibitem{Colangelo:2001df}
  G.~Colangelo, J.~Gasser and H.~Leutwyler,
  Nucl.\ Phys.\ B {\bf 603}, 125 (2001)

\bibitem{inprep}
  R.~Kaminski, J.~R.~Pelaez and F.~J.~Yndurain, in preparation.


\end{thebibliography}
\end{document}